\documentclass[%
 reprint,
 amsmath,amssymb,
 aps,
]{revtex4-2}
\usepackage{amsmath}
\usepackage{amssymb}
\usepackage{graphicx}
\usepackage{dcolumn}
\usepackage{bm}

\begin{document}

\preprint{APS/123-QED}

\title{Intrinsic Angular Momentum Contribution to the Sagnac Response of a Rotating Helical Optical Fiber}

\author{Lili Yang$^1$}
\email{yanglli7@mail.sysu.edu.cn}
\author{Pengming Zhang$^1$}
\author{Longlong Feng$^1$}
\affiliation{
$^1$School of Physics and Astronomy, Sun Yat-Sen University, Zhuhai 519082, China}
\date{\today}

\begin{abstract}
We investigate rotation-induced phase accumulation of structured light
in a mechanically rotating uniform helical optical fiber using a
four-component optical Dirac-like equation. In the paraxial and
weak-geometry regime, the intrinsic dynamics contain a torsion-induced
geometric coupling and a mechanical rotation--angular-momentum
coupling. We show that, in a common-path mirror-reflection geometry,
both the torsion-induced geometric phase and the conventional
area-dependent Sagnac contribution cancel over a complete round trip,
whereas the intrinsic rotation--angular-momentum phase survives. The
surviving phase depends on the intrinsic total angular momentum of the
structured optical mode and provides a common-path interferometric
route for probing intrinsic angular momentum effects in rotational
sensing.
\end{abstract}

\maketitle

\section{Introduction}

The conventional Sagnac effect is a rotation-induced spacetime effect
and forms the fundamental operating principle of fiber-optic
gyroscopes. For two counterpropagating optical fields, the resulting
phase difference is determined by the oriented area enclosed by their
trajectories and may be associated with the coupling of the mechanical
rotation to the extrinsic orbital angular momentum(OAM) of the optical path
\cite{Post1967,Anandan1981,Senthilkumaran2000,Lefevre2014}.
Structured light, however, also carries intrinsic angular momentum (AM), including spin angular momentum (SAM), primarily associated with polarization, and
intrinsic OAM, associated with the spatial
structure of the optical field. In particular, vortex beams with an azimuthal phase
dependence $e^{il\phi}$ carry a well-defined intrinsic OAM characterized
by the integer topological charge $l$
\cite{Allen1992,AllenPadgettBabiker1999,YaoPadgett2011,
BarnettBabikerPadgett2017}. 

When the propagation direction or the optical environment varies in space, the intrinsic AM of light
can become coupled to its trajectory. This interplay underlies a broad class of spin--orbit interaction
phenomena, including the optical Magnus effect, the spin Hall effect of
light, and OAM-dependent geometric transport
\cite{LibermanZeldovich1992,BliokhBliokh2004,Onoda2004,
Bliokh2006,HostenKwiat2008,Bliokh2008GeometrodynamicsOS,
Bliokh2008Coriolis,Bliokh2009a,BliokhBekshaevNori2013,
BliokhDresselNori2014,BliokhNori2015,BliokhEtAl2015SOI,
yang2025inducedberryconnectionphotonic,
yang2025spinorbitcouplinghelicalwaveguides}.
In curved and twisted fibers, this geometric coupling is naturally
described in a local moving frame, whose evolution follows the changing
propagation direction and generates AM dependent
corrections 
\cite{Bliokh2009a,Alexeyev2009,Yamashita2012}. A distinct but closely related coupling arises from mechanical
rotation, which acts directly on the AM carried by the
optical field. This rotation--AM interaction manifests itself in the
angular or rotational Doppler effect
\cite{Garetz1981,Courtial1998Measurement,Courtial1998Rotational,
Mashhoon2002,HauckMashhoon2003}. and the role of optical OAM in
rotating systems and rotational frequency shifts has been demonstrated
both theoretically and experimentally with structured light 
\cite{Speirits2014,Lavery2014}. More generally, such effects are described by the rotation--AM interaction $\boldsymbol{\Omega}\cdot\hat{\mathbf J}$. The AM entering this interaction, however, contains
physically distinct contributions. The extrinsic OAM is associated with
the overall optical trajectory and underlies the conventional
area-dependent Sagnac response, whereas the intrinsic part, 
$\hat{\mathbf J}_{\rm int}
=\hat{\mathbf L}_{\rm int}+\hat{\mathbf S}$,
is carried by the internal spatial and polarization structure of the
guided mode. In a rotating helical waveguide, where both the optical
trajectory and the local propagation frame evolve, separating these
extrinsic and intrinsic rotation-induced contributions becomes
particularly important.

A convenient framework for separating these effects combines
transformation optics with the Dirac-like formulation of Maxwell's
equations. Transformation optics maps curved-space electrodynamics onto
an effective medium with geometry-dependent constitutive tensors
\cite{Pendry2006,LeonhardtPhilbin2009}, while the wave-function
formulation provides a compact operator description of optical angular
momentum and its coupling to geometric and mechanical degrees of
freedom\cite{BialynickiBirula1996,KELLER20051,Barnett2014,
Feng2022FourvectorOD,yang2025inducedberryconnectionphotonic,
YANG2026170557,yang2026photonsmediasecondquantizationscheme}.
Together, these approaches allow the torsion-induced and
mechanical-rotation-induced AM couplings in a helical
waveguide to be identified within a unified framework.

In this work, we develop such a description for structured light
propagating in a mechanically rotating uniform helical optical fiber.
Starting from the effective constitutive relations in the local Frenet
frame, we derive a four-component optical Dirac-like equation and show
that, in the leading paraxial and weak-geometry limit, the two dominant
intrinsic AM couplings are
$\gamma^0\tau\hat J_s$ and $-\Omega_s\hat J_s$.
Here $\tau$ is the torsion of the helical trajectory,
$\Omega_s$ is the longitudinal projection of the mechanical rotation,
and $\hat J_s$ is the intrinsic total AM along the local
propagation direction. The torsion-induced term is branch dependent and
gives rise to a geometric phase, whereas the mechanical-rotation term
is branch independent and describes the direct coupling between
mechanical rotation and the intrinsic AM of light.

To exploit this distinction, we consider a common-path
mirror-reflection configuration in which the optical field propagates
through the helical fiber, is reflected at the end mirror, and retraces
the same trajectory in the opposite direction. We show that the
torsion-induced geometric phase is reciprocal and therefore cancels
over the complete round trip. The conventional area-dependent Sagnac
contribution likewise vanishes for the retraced path. In contrast, the phase arising from the intrinsic AM--rotation coupling survives the round trip. This isolates a net intrinsic AM contribution to the rotation-induced phase, which may be regarded as an AM correction to the conventional Sagnac response.

\section{Dirac-Like Optical Dynamics in a Rotating Helical Fiber}
\label{sec:rotation_coupling}

We consider light propagating in a uniform helical optical fiber with
constant curvature $\kappa$ and torsion $\tau$, while the entire fiber
undergoes a rigid rotation about the global $z$ axis with angular
velocity $\boldsymbol{\Omega}= \Omega_z\,\mathbf e_z $, as shown in
Fig.~\ref{fig:common_path_geometry}. Within the effective-medium description of the curvilinear rotating
system, the mechanical rotation gives rise to a magnetoelectric
coupling. The constitutive relations can be written as\cite{1989Optical,osti_7281940,Leonhardt_2006,Plebanski1960,ThompsonCummerFrauendiener2011,ThompsonFrauendiener2010}
\begin{equation}
\begin{aligned}
\label{eq:constitutive_main}
    \mathbf D
    =
    \boldsymbol{\varepsilon}\mathbf E
    -
    \mathbf g_\Omega\times\mathbf H,
    \qquad
    \mathbf B
    =
    \boldsymbol{\mu}\mathbf H
    +
   g_\Omega\times \mathbf E,
    \end{aligned}
\end{equation}
where $\mathbf g_\Omega= \boldsymbol{\Omega}\times\mathbf R .$ For later convenience, we introduce the cross-product operator $ \mathsf C_\Omega\mathbf V \equiv \mathbf g_\Omega\times\mathbf V$. In terms of this operator, Eq.~\eqref{eq:constitutive_main} can be
inverted perturbatively. To first order in the mechanical rotation,
one obtains
\begin{equation}
\begin{aligned}
         \mathbf E
    &\simeq
    \boldsymbol{\varepsilon}^{-1}\mathbf D
    +
    \boldsymbol{\varepsilon}^{-1}
    \mathsf C_\Omega
    \boldsymbol{\mu}^{-1}\mathbf B,
    \\
    \mathbf H
    &\simeq
    \boldsymbol{\mu}^{-1}\mathbf B
    -
    \boldsymbol{\mu}^{-1}
    \mathsf C_\Omega
    \boldsymbol{\varepsilon}^{-1}\mathbf D .
\end{aligned}
\end{equation}
The inverse constitutive tensors are determined by the effective metric of the helical waveguide. We introduce the local Frenet coordinates $(x_1,x_2,s)$, where $s$ is the arclength along the helical centerline, while $x_1$ $x_2$ denote the transverse coordinates in the local cross section. It is convenient to further introduce the circular transverse coordinates $r_\pm=(x_1\mp i x_2)/\sqrt{2}$. In the corresponding helicity basis, the inverse constitutive tensors take the form\cite{Takagi1992QuantumMO,yang2025inducedberryconnectionphotonic}
\begin{equation}
\label{eq:G_hel_main}
\boldsymbol{\varepsilon}^{-1}
=
\boldsymbol{\mu}^{-1}
\equiv
\mathsf G_{\mathrm{hel}}
=
\begin{pmatrix}
1
&
0
&
-i\tau r_+
\\[2mm]
0
&
1
&
i\tau r_-
\\[2mm]
i\tau r_-
&
-i\tau r_+
&
G_{ss}
\end{pmatrix},
\end{equation}
where
\begin{equation}
    G_{ss}
    =
    h^2+2\tau^2r_+r_-,
    \qquad
    h
    =
    1-\frac{\kappa}{\sqrt{2}}(r_++r_-).
\end{equation}
Let $\mathsf C^{\rm hel}_\Omega$ denote the helicity-basis
representation of $\mathsf C_\Omega$. The corresponding effective
constitutive contribution then reads
\begin{equation}
\label{eq:Momega_main}
\mathsf M_\Omega
=\mathsf G_{\mathrm{hel}}
    \mathsf C^{hel}_\Omega
    \mathsf G_{\mathrm{hel}}=
\begin{pmatrix}
g_0
&
0
&
iKr_+
\\[2mm]
0
&
-g_0
&
iKr_-
\\[2mm]
-iKr_-
&
-iKr_+
&
0
\end{pmatrix},
\end{equation}
with
\begin{equation}
\label{eq:g0_main}
    g_0
    =
    g_{\Omega s}
    -
    2\tau\Omega_s r_+r_-,\, K
    =
    \Omega_sG_{ss}
    -
    \tau g_{\Omega s}.
\end{equation}
To expose the Dirac-like structure, we introduce the four-component
optical wave function\cite{Feng2022FourvectorOD,yang2025inducedberryconnectionphotonic}
\begin{equation}
    \boldsymbol{\Psi}
    =
    \begin{pmatrix}
        \boldsymbol{\Psi}_+
        \\
        \boldsymbol{\Psi}_-
    \end{pmatrix},
    \qquad
    \boldsymbol{\Psi}_\pm
    =
    \mathbf D_\perp
    \pm
    i\sigma_3\mathbf B_\perp .
\end{equation}
Following the derivation presented in Appendix~\ref{appendix1}, the
propagation of light in a rotating helical fiber is governed by the
optical Dirac-like equation
\begin{equation}
\label{eq:effective_main}
i\frac{\partial\boldsymbol{\Psi}}{\partial t}
+
\Omega_s\hat J_s\boldsymbol{\Psi}
=
\left[
\gamma^0\hat k_s
+
\gamma^0\tau\hat J_s
\right]
\boldsymbol{\Psi}
+
\hat H'\boldsymbol{\Psi},
\end{equation}
where $\hat k_s=-i\partial_s$ is the longitudinal momentum operator
along the local propagation direction. 
Equation~\eqref{eq:effective_main} makes explicit the two leading
AM couplings,
\begin{equation}
\label{eq:rotation_torsion_terms}
    \hat{H}_{\tau}
    =
    \gamma^0\tau\hat{J}_s,
    \qquad
    \hat{H}_{\Omega}
    =
    -\Omega_s\hat{J}_s .
\end{equation}
The torsion-induced term $\gamma^0\tau\hat J_s$ gives rise to the
geometric phase accumulated during propagation along the helical
trajectory
\cite{Chiao1986ManifestationsOB,Tomita1986Berry,Bliokh2006},
whereas the mechanical-rotation term $-\Omega_s\hat J_s$ represents
the rotation--AM coupling underlying the rotational
Doppler effect
\cite{Garetz1981,1989Optical,Courtial1998Measurement,
Courtial1998Rotational,Mashhoon2002,HauckMashhoon2003,
YANG2026170557}. All remaining terms, including the
correction $\delta K$, transverse mode-mixing contributions,
higher-order paraxial terms, and operator-ordering corrections, are
higher-order geometric effects and are collected into $\hat H'$. These
terms can be treated perturbatively under the above weak-geometry
conditions, and their explicit expressions are provided in
Appendix~\ref{appendix1}.

The spin and total AM operators along the local
propagation direction are defined as
\begin{equation}
\label{eq:Ss_Js}
    \hat S_s
    =
    \begin{pmatrix}
    \sigma_3&0
    \\
    0&\sigma_3
    \end{pmatrix},
    \qquad
    \hat J_s
    =
    \hat L_s\mathbf 1_4+\hat S_s .
\end{equation}
The two components $\boldsymbol{\Psi}_+$ and
$\boldsymbol{\Psi}_-$ define the two eigensectors of the branch
operator $\gamma^0$,
\begin{equation}
        \gamma^0\boldsymbol{\Psi}_\eta
    =
    \eta\boldsymbol{\Psi}_\eta,
    \qquad
    \eta=\pm1 .
\end{equation}
In the leading paraxial limit,
$i\partial_t\boldsymbol{\Psi}
\simeq\gamma^0\hat k_s\boldsymbol{\Psi}$,
which gives the dispersion relation
$\omega=\eta k_s$ , where $k_s$ here denotes the
signed longitudinal wave vector. For a fixed sign of $k_s$, the two
$\gamma^0$ sectors therefore correspond formally to the positive- and
negative-frequency branches\cite{KELLER20051}.

For the physical electromagnetic field, however, we restrict attention
to the positive-frequency sector, $\omega>0$. Reversing the propagation
direction, $k_s\rightarrow-k_s$ must then be accompanied by
$\eta\rightarrow-\eta$, so that the product $\eta k_s$, and hence
$\omega$, remains positive. Accordingly, if $k_s>0$ is now used to
denote the magnitude of the longitudinal wave vector, the forward- and
backward-propagating solutions can be written as
\begin{equation}
\begin{aligned}
    \boldsymbol{\Psi}_{+}
\propto
    \boldsymbol{\Psi}_{0,+}
    e^{-i\omega t+i k_s s},\,
    \boldsymbol{\Psi}_{-}\propto
    \boldsymbol{\Psi}_{0,-}
    e^{-i\omega t-i k_s s},
\end{aligned}
\end{equation}
where $\boldsymbol{\Psi}_{0,\pm}$ are constant field amplitudes.
Both solutions correspond to the same positive physical frequency,
with $\omega=k_s$. Thus, within the
positive-frequency sector, the $\eta=+1$ and $\eta=-1$ eigensectors of
$\gamma^0$ represent the forward- and backward-propagating modes,
respectively.

\section{Intrinsic Angular-Momentum-Induced Rotation Phase in a Common-Path Geometry}
\label{sec:rotation_phase}

\begin{figure*}[t]
    \centering
    \includegraphics[width=\linewidth]{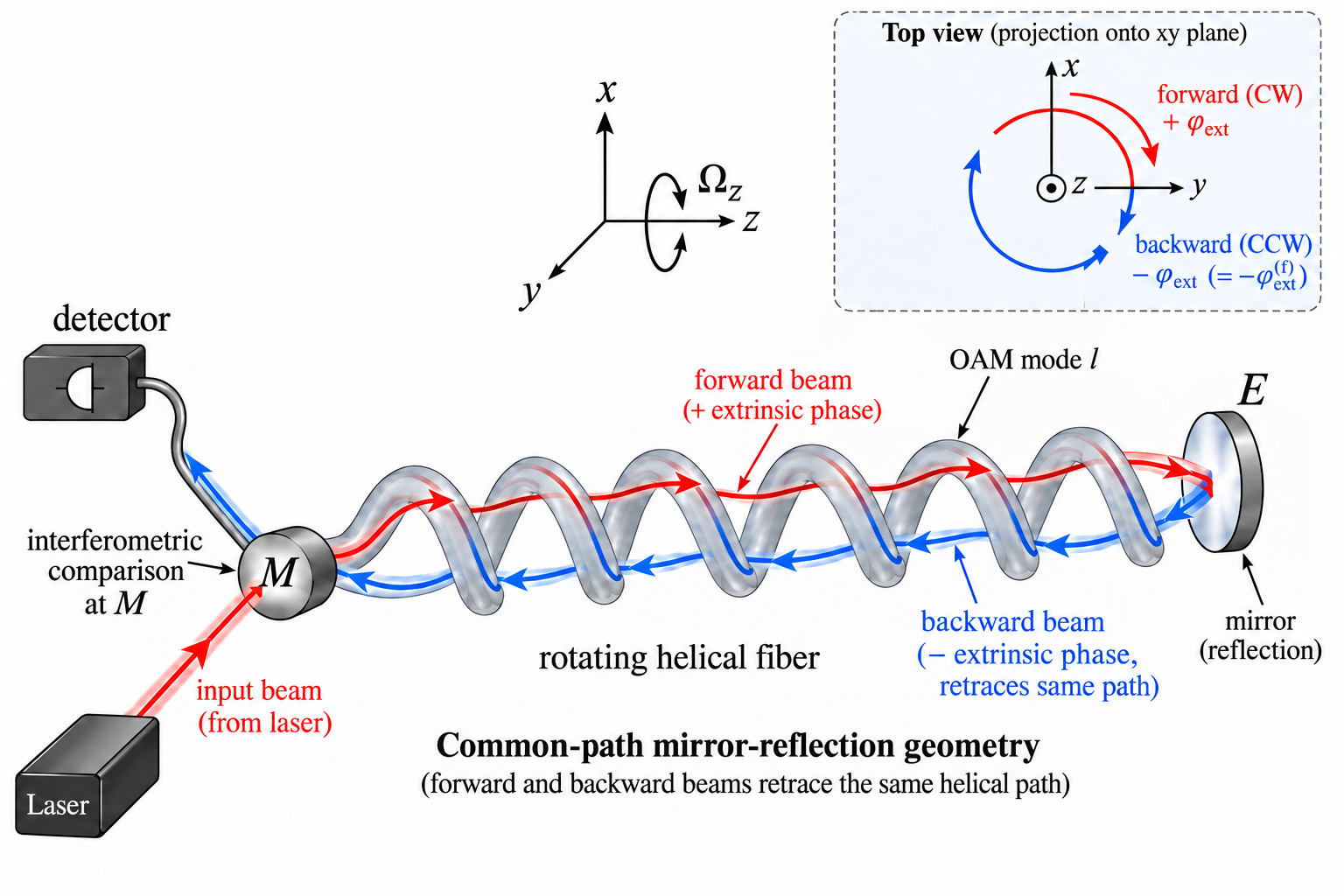}
    \caption{
Schematic of the common-path mirror-reflection configuration in a
    rotating helical fiber. A vortex mode propagates from $M$ to the end mirror $E$ and returns along the
    same helical trajectory after reflection. The forward and backward
    beams are indicated by the red and blue paths, respectively, and are
    interferometrically compared at $M$. The inset shows the $xy$-plane projection of the helical trajectory,
for which the forward and backward paths circulate in opposite
directions about the $z$ axis.
    }
    \label{fig:common_path_geometry}
\end{figure*}

We now turn to the intrinsic AM-dependent rotation phase.
To isolate this contribution, we consider the common-path
mirror-reflection geometry shown in Fig.~\ref{fig:common_path_geometry}. Light propagates from one end of the helical fiber, denoted by $M$, to
the opposite end $E$, where it is reflected by a mirror and subsequently
retraces the same helical path back to $M$. The returned field is then
interferometrically compared at $M$. Although the two passages follow
the same spatial trajectory, their projections onto the plane
perpendicular to the rotation axis correspond to opposite azimuthal
orientations with respect to
$\boldsymbol{\Omega}$. Therefore, each propagation segment acquires an extrinsic
rotation-induced phase analogous to the Sagnac phase, with opposite
signs for the forward and backward passages.

More generally, in a rotating reference frame the AM-rotation
coupling can be written as
\begin{equation}
\label{eq:full_rotation_coupling}
    \hat{H}_{\Omega}^{\rm tot}
    =
    -
    \boldsymbol{\Omega}\cdot
    \left(
    \hat{\mathbf L}_{\rm ext}
    +
    \hat{\mathbf L}_{\rm int}
    +
    \hat{\mathbf S}
    \right),
\end{equation}
where $\hat{\mathbf L}_{\rm ext}$ denotes the extrinsic orbital angular
momentum associated with the motion of the center of the optical path,
whereas $\hat{\mathbf L}_{\rm int}$ and $\hat{\mathbf S}$ denote,
respectively, the intrinsic orbital and spin angular momenta of the
guided optical mode. The conventional area-dependent Sagnac phase
originates from the extrinsic contribution
$-\boldsymbol{\Omega}\cdot\hat{\mathbf L}_{\rm ext}$. In the present mirror-reflection geometry, the forward $(+)$
and backward $(-)$ passages are two successive propagation
segments of the same optical field. The total extrinsic phase
accumulated over the complete round trip is
$\Phi_{\rm ext}^{(\mathrm{rt})}
=\phi_{\rm ext}^{(+)}+\phi_{\rm ext}^{(-)}$.
Since the backward trajectory exactly retraces the forward
path in the opposite direction, the corresponding line
integrals have opposite orientations, yielding
$\phi_{\rm ext}^{(-)}=-\phi_{\rm ext}^{(+)}$.
Consequently, the extrinsic contribution vanishes,
$\Phi_{\rm ext}^{(\mathrm{rt})}=0$,
as follows from the line-integral expression in
Appendix~\ref{appendix3}.
As shown below, the torsion-induced geometric phase
also cancels over the retraced path, leaving the
intrinsic AM-rotation coupling as the remaining
rotation-dependent contribution within the present
approximation.

We next demonstrate how the torsion-induced geometric
phase cancels in the mirror-reflection geometry.
To describe both propagation directions consistently,
we adopt the same fixed Frenet frame.
Under ideal, mode-preserving normal reflection, the
reflection operator satisfies
\begin{equation}
    \hat R_{\rm mirror}^{-1}
    \gamma^0
    \hat R_{\rm mirror}
    =
    -\gamma^0,
    \qquad
    \hat R_{\rm mirror}^{-1}
    \hat J_s
    \hat R_{\rm mirror}
    =
    \hat J_s.
\end{equation}
As derived in Appendix~\ref{app:mirror_reflection},
reflection exchanges the two propagation branches
while preserving the intrinsic AM eigenvalue
$j=l+\sigma$ along the fixed Frenet axis.
When referred to the actual propagation direction,
however, the OAM and SAM
quantum numbers both reverse sign.

Neglecting the mechanical-rotation term and the higher-order
corrections, Eq.~(\ref{eq:effective_main}) reduces to
\begin{equation}
    i\frac{\partial\boldsymbol{\Psi}}{\partial t}
    =
    \gamma^0
    \left(
        \hat k_s+\tau\hat J_s
    \right)
    \boldsymbol{\Psi}.
\end{equation}
For the forward-propagating branch $\boldsymbol{\Psi}_+$,
which is an eigenstate of $\gamma^0$ with eigenvalue $+1$,
the corresponding dispersion relation is
$\omega=k_++\tau j$, where $j=l+\sigma$ is the
eigenvalue of the intrinsic AM along the fixed Frenet axis.
For a monochromatic field with fixed physical frequency
$\omega$, the corresponding longitudinal propagation constant
is therefore $k_+=k_0-\tau j$, where $k_0=\omega$.
The field accumulated over a propagation length $L$ is
consequently
\begin{equation}
    \boldsymbol{\Psi}_+(L,t)
    \propto
    e^{-i\omega t}
    e^{ik_0L}
    e^{-i\tau jL}.
\end{equation}
Thus, the torsion-induced geometric phase acquired during
the forward passage is
\begin{equation}
    \phi_\tau^{(+)}
    =
    -\tau jL.
\end{equation}

After reflection at the end mirror, the propagation direction
is reversed and the field is transferred to the opposite
branch $\boldsymbol{\Psi}_-$, for which $\gamma^0$ has
eigenvalue $-1$. Importantly, the intrinsic AM eigenvalue
$j$ remains unchanged because both branches are described
in the same fixed Frenet frame. Denoting by $k_->0$ the
magnitude of the backward longitudinal wave vector, the
eigenvalue of $\hat k_s$ in this frame is $-k_-$.
The corresponding dispersion relation is therefore
$\omega=-(-k_-+\tau j)=k_--\tau j$.
For the same fixed frequency $\omega$, this gives
$k_-=k_0+\tau j$. Hence, during the backward propagation
over the same physical length $L$, the torsion-induced
phase is
\begin{equation}
    \phi_\tau^{(-)}
    =
    +\tau jL.
\end{equation}
The geometric phase accumulated over the complete round
trip therefore vanishes,
\begin{equation}
\label{eq:tau_roundtrip_cancel}
    \Phi_\tau^{(\mathrm{rt})}
    =
    \phi_\tau^{(+)}
    +
    \phi_\tau^{(-)}
    =
    0.
\end{equation}
Such a cancellation of the geometric phase upon reflection
and retracing of a helically wound fiber has also been
demonstrated experimentally \cite{Senthilkumaran2000}.

We now consider the intrinsic AM-dependent phase arising
from mechanical rotation. Unlike the torsion-induced
coupling $\gamma^0\tau\hat J_s$, the mechanical-rotation
coupling $-\Omega_s\hat J_s$ does not contain the branch
operator $\gamma^0$. Since the intrinsic AM eigenvalue
$j$ is preserved under reflection in the fixed Frenet
frame, the forward-propagating mode acquires the phase
\begin{equation}
    \phi_{\mathrm{int}}^{(+)}
    =
    +\Omega_s jT.
\end{equation}
After reflection, the rotation-induced contribution
retains the same sign,
\begin{equation}
    \phi_{\mathrm{int}}^{(-)}
    =
    +\Omega_s jT.
\end{equation}
The mechanical-rotation phase therefore accumulates
constructively over the complete round trip,
\begin{equation}
\label{eq:Omega_roundtrip}
    \Phi_{\mathrm{int}}^{(\mathrm{rt})}
    =
    \phi_{\mathrm{int}}^{(+)}
    +
    \phi_{\mathrm{int}}^{(-)}
    =
    2\Omega_s jT.
\end{equation}

The above analysis reveals a clear distinction between
the torsion-induced geometric phase and the phase
generated by mechanical rotation. Their different
behaviors under reflection originate from the presence
or absence of the branch operator $\gamma^0$.
In the fixed Frenet representation, mirror reflection
reverses the propagation branch while preserving the
intrinsic AM eigenvalue $j$. Consequently, the
torsion-induced geometric phases cancel, whereas the
intrinsic rotation-induced phases retain the same sign
and add constructively. The behavior of the relevant
phase contributions under propagation reversal is
summarized in Table~\ref{tab:phase_summary}.

\begin{table}[t]
\centering
\caption{
Phase contributions for forward and backward propagation in the
common-path mirror-reflection geometry.
}
\label{tab:phase_summary}
\begin{tabular}{lccc}
\hline\hline
Interaction
& Forward
& Backward
& Round trip
\\
\hline
$\gamma^0\tau\hat J_s$
& $-\tau jL$
& $+\tau jL$
& $0$
\\[1mm]

$-\Omega_s\hat J_s$
& $+\Omega_s jT$
& $+\Omega_s jT$
& $2\Omega_s jT$
\\[1mm]

$-\boldsymbol{\Omega}\!\cdot\!\hat{\mathbf L}_{\rm ext}$
& $+\phi_{\rm Sag}^{(\mathrm +)}$
& $-\phi_{\rm Sag}^{(\mathrm -)}$
& $0$
\\
\hline\hline
\end{tabular}
\end{table}
The last two rows originate from the same fundamental
rotation--AM coupling,
$-\boldsymbol{\Omega}\cdot\hat{\mathbf J}$, but represent its intrinsic
and extrinsic sectors, respectively. The conventional Sagnac phase is
associated with the extrinsic OAM
$\hat{\mathbf L}_{\rm ext}$ of the optical trajectory. Reversing the
propagation direction reverses the orientation of the corresponding
line integral, so that the extrinsic Sagnac phase changes sign and
cancels over the retraced round trip. By contrast, the intrinsic total
AM, $(\hat{\mathbf L}_{\rm int}
    +
    \hat{\mathbf S})$, couples directly to the mechanical rotation and produces phase
contributions of the same sign during the forward and backward
passages. The intrinsic rotation-induced phase therefore remains
nonzero after the complete round trip. The torsion-induced
interaction $\gamma^0\tau\hat J_s$ is likewise reciprocal, but for a
different reason: it explicitly contains the branch operator
$\gamma^0$ and hence changes sign between the forward- and
backward-propagating sectors. Consequently, both the conventional area-dependent
Sagnac phase and the torsion-induced geometric phase vanish in the
retraced common-path geometry, whereas the intrinsic
rotation--AM phase survives. The mirror-reflection
configuration thus provides a natural means of isolating the
rotation-induced phase associated with the intrinsic total angular
momentum of the guided optical mode.

For a uniform helix, let $q=\sqrt{\kappa^2+\tau^2}$. The tangential
projection of the global rotation is $\Omega_s=\Omega_z\tau/q$, and
the length and group delay over $N$ turns are $L_N=2\pi N/q$ and
$T_{N,j}=L_N/v_{g,j}$, respectively. In the leading-order,
mode-preserving approximation, the intrinsic round-trip phase is
\begin{equation}
\label{eq:intrinsic_roundtrip_pitch}
\Phi_{\mathrm{int,prop}}^{(\mathrm{rt})}
\simeq 2\Omega_s jT_{N,j}
=\frac{4\pi N\Omega_z p_h}{v_{g,j}}\,j,
\quad
p_h=\frac{\tau}{\kappa^2+\tau^2},
\end{equation}
where $2\pi p_h$ is the axial pitch per turn and $v_{g,j}$ is the
modal group velocity.

We consider a hollow perfect-electric-conductor (PEC) helical waveguide
terminated by a PEC mirror. The helical section comprises one turn
($N=1$), with $L_N=3.9738353\,\mathrm{mm}$, and is connected to a
stationary $4\,\mu\mathrm{m}$ input lead. The three-dimensional waveguide
geometry and PEC boundary configuration are modeled using COMSOL
Multiphysics. The modal scattering response is calculated by solving
the vector Maxwell equations through spectral mode matching in the
exact Frenet-coordinate metric of the helical geometry. The intrinsic
rotational correction is retained to first order through
$\mathbf{g}_{\mathrm{int}}
=\boldsymbol{\Omega}\times\boldsymbol{\xi}_{\perp}$.
The transverse basis contains 340 TE/TM channels, including
83 propagating and 257 evanescent channels.

\begin{figure}[t]
\centering
\includegraphics[width=0.98\columnwidth]{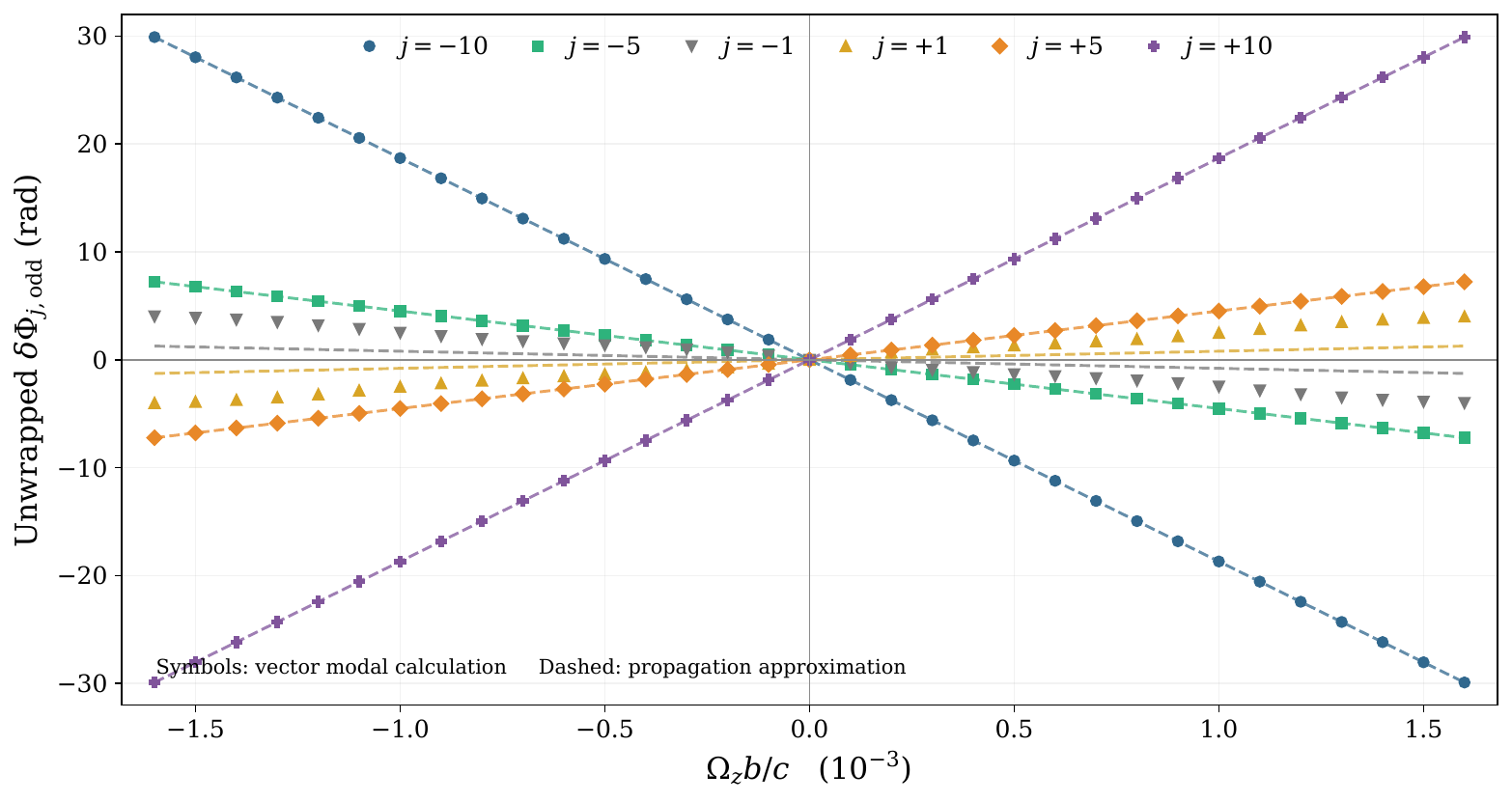}
\caption{Rotation-odd modal return phases of a one-turn hollow
PEC helical waveguide terminated by a PEC mirror.
Symbols show vector-Maxwell spectral mode-matching results for
$\mathrm{TE}_{|j|,1}$ input modes with $j=\pm1,\pm5,\pm10$;
dashed curves represent the leading mode-preserving prediction
in Eq.~\eqref{eq:intrinsic_roundtrip_pitch}.
The 129 rotation samples span
$-1.6\times10^{-3}\leq\Omega_z b/c\leq1.6\times10^{-3}$,
with every fourth sample displayed.
Parameters are $b=3.2\,\mu\mathrm{m}$,
$R=600\,\mu\mathrm{m}$, $p_h=200\,\mu\mathrm{m}$,
$\lambda=1.55\,\mu\mathrm{m}$,
$\varepsilon_r=\mu_r=1$, and $N=1$.
A stationary input lead of length $4\,\mu\mathrm{m}$ is included.
The spectral basis contains 83 propagating and
257 evanescent TE/TM channels, and only the first-order
intrinsic rotation perturbation is retained.}
\label{fig:oam_rotation_phase}
\end{figure}

The rotation-induced same-mode return phase is defined relative to
zero rotation by
\begin{equation}
\label{eq:same_mode_return_phase}
\widetilde{\Delta\Phi}_j(\Omega_z)
=\operatorname{unwrap}\!\left\{
\arg\!\left[
r_{jj}(\Omega_z)r_{jj}^{*}(0)
\right]
\right\},
\quad
\widetilde{\Delta\Phi}_j(0)=0.
\end{equation}
Here $r_{jj}$ is the complex reflection coefficient from the incident
mode $j$ into the corresponding backward-propagating mode at the
input port. The phase is continuously unwrapped along the sampled
rotation-rate sweep starting from $\Omega_z=0$, thereby removing
the common static propagation and mirror-reflection phases.
The rotation-odd component is then extracted as
\begin{equation}
\label{eq:rotation_odd_phase}
\delta\Phi_{j,\mathrm{odd}}
=\frac{
\widetilde{\Delta\Phi}_j(+\Omega_z)
-\widetilde{\Delta\Phi}_j(-\Omega_z)
}{2}.
\end{equation}
As shown in Fig.~\ref{fig:oam_rotation_phase}, the rotation-odd
phases reverse sign for opposite total angular momenta.
At $\Omega_z b/c=1.6\times10^{-3}$, the numerical results are
\begin{equation}
\label{eq:rotation_odd_numeric}
\begin{aligned}
\delta\Phi_{\pm1,\mathrm{odd}}
&\simeq\pm4.00847\,\mathrm{rad},\\
\delta\Phi_{\pm5,\mathrm{odd}}
&\simeq\pm7.22673\,\mathrm{rad},\\
\delta\Phi_{\pm10,\mathrm{odd}}
&\simeq\pm29.90412\,\mathrm{rad}.
\end{aligned}
\end{equation}
The corresponding mode-preserving estimates are $\pm1.26949$,
$\pm7.22928$, and $\pm29.90466\,\mathrm{rad}$, respectively.
For $|j|=5$ and 10, the relative deviations are approximately
$0.0353\%$ and $0.00181\%$, demonstrating close agreement with
the leading-order single-mode propagation model.

A substantially larger discrepancy occurs for $|j|=1$.
At zero rotation, only $9.17\%$ of the incident power is
reflected into the original mode, indicating strong depletion
of the same-mode return channel. The projected reflection
amplitude $r_{jj}$ may therefore contain interfering contributions
from multiple propagation and mode-conversion pathways.
Consequently, its phase need not coincide with the propagation
phase of an isolated mode and may become particularly sensitive
to rotation under destructive interference. At the benchmark
rotation rate, the deviation from the mode-preserving estimate
reaches approximately $2.74\,\mathrm{rad}$.
Increasing the transverse basis from 239 to 340 channels changes
the extracted rotation-odd phases by no more than
$1.39\times10^{-6}\,\mathrm{rad}$ over the sampled rotation rates.
The large discrepancy for $|j|=1$ is therefore consistent with
a breakdown of the mode-preserving approximation rather than
insufficient transverse-basis convergence. A definitive
identification of the dominant mode-conversion pathways would
require a mode-resolved analysis of the scattering matrix.

At the detection point $M$, the total optical field is
$\boldsymbol{\Psi}_{\rm det}
=\boldsymbol{\Psi}_{\rm ref}+\boldsymbol{\Psi}_{\rm rt}$,
where “ref" and "rt" denote the reference and
round-trip returned fields, respectively.
The detected optical intensity is proportional
to the quadratic norm of the optical wave function,
\begin{equation}
\begin{aligned}
I_{\rm det}
=
\mathcal C\,
\boldsymbol{\Psi}_{\rm det}^{\dagger}
\boldsymbol{\Psi}_{\rm det}
=
I_{\rm ref}+I_{\rm rt}
+2\mathcal C\,
\operatorname{Re}
\left(
\boldsymbol{\Psi}_{\rm ref}^{\dagger}
\boldsymbol{\Psi}_{\rm rt}
\right),
\end{aligned}
\end{equation}
where $\mathcal C$ is a normalization factor
determined by the modal normalization and
detection scheme. Assuming that the reference and returned fields
are mode matched at $M$, their relative phase
can be written as
$\phi_{\rm rt}-\phi_{\rm ref}
=\Phi_{\rm int}^{(\mathrm{rt})}+\phi_0$,
where $\Phi_{\rm int}^{(\mathrm{rt})}$ is the intrinsic
rotation-induced round-trip phase and $\phi_0$
denotes the static background phase of the
interferometer.
Introducing the background intensity
$I_0=I_{\rm ref}+I_{\rm rt}$ and the interference
visibility
$V=2\sqrt{I_{\rm ref}I_{\rm rt}}/I_0$
for perfectly mode-matched fields, the normalized
intensity variation at the quadrature point
$\phi_0=\pi/2$ becomes
\begin{equation}
\begin{aligned}
\frac{\delta I_{\rm det}}{I_0}\simeq
-V\Phi_{\rm int}^{(\mathrm{rt})},
\end{aligned}
\end{equation}
where
$\delta I_{\rm det}
=I_{\rm det}(\Phi_{\rm int}^{(\mathrm{rt})})
-I_{\rm det}(0)$
and $|\Phi_{\rm int}^{(\mathrm{rt})}|\ll1$.

Previous Sagnac-based studies have demonstrated
OAM-enhanced rotation-vibration measurements
using a rotating Dove prism \cite{Xiao2018}
and the interferometric characterization of
vortex beams carrying different OAM states
\cite{Aili2022}.
In contrast, the rotation-induced phase
considered here arises from the coupling
between the mechanical rotation of a helical
waveguide and the intrinsic total angular
momentum $j=l+\sigma$ of the guided field.
This mechanism provides a possible route
toward rotation sensing that exploits both
the OAM and helicity degrees of freedom,
without requiring a rotating Dove prism. In particular, the contributions of orbital
and spin angular momentum could, in principle,
be distinguished by comparing the responses
of modes with opposite topological charges
or opposite helicities.
For otherwise identical propagation conditions,
the corresponding phase differences are
\begin{equation}
\begin{aligned}
\Phi_{l,\sigma}-\Phi_{-l,\sigma}
&=
\frac{8\pi N\Omega_zp_h}{v_g}\,l,\\
\Phi_{l,+}-\Phi_{l,-}
&=
\frac{8\pi N\Omega_zp_h}{v_g}.
\end{aligned}
\end{equation}

For comparison with the conventional Sagnac interferometer discussed in
Eq.~(\ref{eq:Sagnac_external}), the corresponding area-dependent phase is 
\begin{equation}
    \Delta\Phi_{\rm Sag}^{(0)}
    =
    \frac{4\pi N\omega\Omega_zR^2}{c^2}.
\end{equation}
Although the conventional area-dependent Sagnac contribution vanishes
in the retraced common-path geometry, it provides a useful reference
scale for assessing the relative magnitude of the OAM-dependent
rotation phase. Retaining the part of
Eq.~\eqref{eq:intrinsic_roundtrip_pitch} proportional to $l$, we define
\begin{equation}
\label{eq:eta_l_new}
    \eta_l
    \equiv
    \frac{\Delta\Phi_l}
         {\Delta\Phi_{\rm Sag}^{(0)}}
    =
    l\,
    \frac{c^2p_h}
         {\omega v_gR^2},
\end{equation}
In particular, $|\eta_l|\gtrsim 1$ identifies
a regime in which the intrinsic OAM-induced
rotation phase becomes comparable to or larger
than the conventional Sagnac phase.
This opens up the possibility of exploiting
high-order vortex modes in compact helical
waveguides for rotation sensing, particularly
in common-path configurations where the
conventional area-dependent contribution
cancels.

\section{Conclusion}

In conclusion, we have investigated the rotation-induced dynamics of
structured light propagating in a mechanically rotating uniform
helical optical fiber. Within an optical Dirac-like framework, we
identified two distinct AM couplings: one originating
from the geometry of the helical trajectory and the other from the
mechanical rotation of the fiber. In the leading paraxial and
weak-geometry regime, they reduce to a torsion-induced geometric
coupling and a mechanical rotation--AM coupling,
respectively.

A key distinction between these two contributions emerges in the
common-path mirror-reflection geometry. The torsion-induced geometric
phase changes sign between the forward and backward propagation
branches and therefore cancels over a complete round trip. The
conventional area-dependent Sagnac phase, which originates from the
coupling of mechanical rotation to the extrinsic orbital motion of the
optical trajectory, is likewise reciprocal for the retraced path and
also cancels. In contrast, the coupling between mechanical rotation and
the intrinsic AM of the guided mode contributes with the
same sign during the forward and backward passages and consequently
survives the round trip. The common-path configuration therefore suppresses the conventional
area-dependent Sagnac background while retaining an
intrinsic AM dependent rotation phase, thereby providing a direct
way to distinguish intrinsic and extrinsic AM
contributions to rotational optical phenomena.

These results extend the conventional Sagnac picture by showing that
mechanical rotation can couple not only to the overall optical
trajectory but also to the internal AM structure of the
guided field. The present framework may provide a useful basis for
exploring rotation sensing with structured light and for studying the
interplay between geometric phases, OAM, SAM, and mechanical rotation in curved photonic systems.

\begin{acknowledgments}
This work was supported by the National Key R\&D Program of China (grant No. 2024YFE0109802), the National Natural Science Foundation of China (grants Nos. 12175320,12375084), the Natural Science Foundation of Guangdong Province, China (Grant No. 2022A1515010280).
\end{acknowledgments}

\appendix

\section{Derivation of the Optical Dirac-Like Equation in a Rotating Helical Fiber}
\label{appendix1}

The curved-coordinate description can be mapped onto an effective
bianisotropic medium with constitutive relations
\begin{equation}
\label{eq:constitutive_rotation}
    \mathbf D
    =
    \boldsymbol{\varepsilon}\mathbf E
    -
    \mathbf g_\Omega\times\mathbf H,
    \qquad
    \mathbf B
    =
    \boldsymbol{\mu}\mathbf H
    +
    \mathbf g_\Omega \times\mathbf E .
\end{equation}

For a rotating system, the rotation-induced vector is generally
$\mathbf g_\Omega=\boldsymbol{\Omega}\times\mathbf R$, where
$\mathbf R$ denotes the spatial position with respect to the rotation
axis. In the local Frenet description, we write
\begin{equation}\label{R}
    \mathbf R
    =
    \mathbf R_c(s)
    +
    \boldsymbol{\xi}_\perp,
    \qquad
    \boldsymbol{\xi}_\perp
    =
    x_1\mathbf e_1+x_2\mathbf e_2 ,
\end{equation}
where $\mathbf R_c(s)$ is the helical centerline and
$\boldsymbol{\xi}_\perp$ is the displacement from the centerline within
the local transverse cross section. For convenience, we introduce the cross-product operator
$\mathsf C_\Omega\mathbf V
\equiv\mathbf g_\Omega\times\mathbf V$.
To first order in the mechanical rotation, the constitutive relations
can be inverted as
\begin{equation}
\begin{aligned}
         \mathbf E
    &\simeq
    \boldsymbol{\varepsilon}^{-1}\mathbf D
    +
    \boldsymbol{\varepsilon}^{-1}
    \mathsf C_\Omega
    \boldsymbol{\mu}^{-1}\mathbf B,
    \\
    \mathbf H
    &\simeq
    \boldsymbol{\mu}^{-1}\mathbf B
    -
    \boldsymbol{\mu}^{-1}
    \mathsf C_\Omega
    \boldsymbol{\varepsilon}^{-1}\mathbf D .
\end{aligned}
\end{equation}
Within the present transformation-optics description,
$\boldsymbol{\varepsilon}^{-1}
=\boldsymbol{\mu}^{-1}\equiv\mathsf G$, so that the
rotation-induced contribution naturally involves the product
$\mathsf G\mathsf C_\Omega\mathsf G$.

We use the local Frenet frame
$\{\mathbf e_1,\mathbf e_2,\mathbf e_s\}$, with $\mathbf e_s$ tangent
to the helical centerline, and decompose the mechanical rotation as
$\boldsymbol{\Omega}
=\Omega_1\mathbf e_1+\Omega_2\mathbf e_2+\Omega_s\mathbf e_s$.
For a uniform helix rotating about the global $z$ axis,
\begin{equation}
\label{eq:Omega_decomposition}
    \Omega_1=0,
    \quad
    \Omega_2
    =
    \frac{\kappa}
    {\sqrt{\kappa^2+\tau^2}}\Omega_z,
    \quad
    \Omega_s
    =
    \frac{\tau}
    {\sqrt{\kappa^2+\tau^2}}\Omega_z .
\end{equation}
In the corresponding local Frenet coordinates $(x_1,x_2,s)$, the
inverse constitutive tensor is\cite{Takagi1992QuantumMO,yang2025inducedberryconnectionphotonic}
\begin{equation}
\label{eq:G_cartesian}
\mathsf G
=
\begin{pmatrix}
1 & 0 & -\tau x_2
\\[2mm]
0 & 1 & \tau x_1
\\[2mm]
-\tau x_2 & \tau x_1
&
h^2+\tau^2(x_1^2+x_2^2)
\end{pmatrix},
\end{equation}
where $h=1-\kappa x_1$. To make the helicity structure explicit, we transform from the local Frenet basis to the circular basis using
\begin{equation}
\label{eq:helicity_transformation}
    U
    =
    \frac{1}{\sqrt{2}}
    \begin{pmatrix}
    1&1&0
    \\
    i&-i&0
    \\
    0&0&\sqrt{2}
    \end{pmatrix}.
\end{equation}
Accordingly, a vector
$\mathbf V=(V_1,V_2,V_s)^T$ is represented as
$\mathbf V_{\rm hel}=U^\dagger\mathbf V=(V_+,V_-,V_s)^T$, with
$V_\pm=(V_1\mp iV_2)/\sqrt{2}$. We similarly introduce the circular
transverse coordinates
$r_\pm=(x_1\mp i x_2)/\sqrt{2}$. In this basis, the inverse constitutive tensor becomes
\begin{equation}
\label{eq:G_hel}
\mathsf G_{\mathrm{hel}}
=
U^\dagger\mathsf G U
=
\begin{pmatrix}
1 & 0 & -i\tau r_+
\\[2mm]
0 & 1 & i\tau r_-
\\[2mm]
i\tau r_- & -i\tau r_+ & G_{ss}
\end{pmatrix},
\end{equation}
where
$G_{ss}=h^2+2\tau^2r_+r_-$. The circular components of the rotation vector are defined by
$\Omega_\pm=(\Omega_1\mp i\Omega_2)/\sqrt{2}$.

Since the centerline contribution
$\boldsymbol{\Omega}\times\mathbf R_c(s)$ is associated with the
overall orbital motion of the fiber and gives rise to the conventional
extrinsic Sagnac response, it is treated separately below. Here we
focus on the local intrinsic sector and, for brevity, define
\begin{equation}
    \mathbf g_\Omega
    \equiv
    \mathbf g_\Omega^{\rm int}
    =
    \boldsymbol{\Omega}\times\boldsymbol{\xi}_\perp .
\end{equation}
The corresponding
helicity components are
\begin{equation}
\begin{aligned}
    (g_\Omega)_+
    &=
    -i\Omega_s r_+,
    \quad
    (g_\Omega)_-
    =
    i\Omega_s r_-,
    \\
    (g_{\Omega})_s
    &=
    -i\left(
        \Omega_+ r_-
        -
        \Omega_- r_+
    \right).
\end{aligned}
\end{equation}
After transforming the cross-product operator to the same helicity
basis, and retaining the notation $\mathsf C_\Omega$ for brevity, the
effective rotation-induced constitutive matrix is
\begin{equation}
\label{eq:Momega_def}
    \mathsf M_\Omega
    =
    \mathsf G_{\mathrm{hel}}
    \mathsf C_\Omega
    \mathsf G_{\mathrm{hel}} .
\end{equation}
Direct multiplication gives
\begin{equation}
\label{eq:Momega}
\mathsf M_\Omega
=
-i
\begin{pmatrix}
g_0 & 0 & iKr_+
\\[2mm]
0 & -g_0 & iKr_-
\\[2mm]
-iKr_- & -iKr_+ & 0
\end{pmatrix},
\end{equation}
with
\begin{equation}
     g_0
    =
    g_{\Omega s}
    -
    2\tau\Omega_s r_+r_-,
    \qquad
    K
    =
    \Omega_sG_{ss}
    -
    \tau g_{\Omega s}.
\end{equation}
For the present axial rotation of a uniform helix, the geometric
relation between the transverse and longitudinal components of
$\boldsymbol{\Omega}$ allows $K$ to be reduced to
\begin{equation}
\label{eq:K_simplified}
K
=
\Omega_s
\left[
1
-\frac{\kappa}{\sqrt{2}}(r_++r_-)
+\frac{\kappa^2}{2}(r_++r_-)^2
+2\tau^2r_+r_-
\right].
\end{equation}
It is therefore useful to separate
$K=\Omega_s+\delta K$, where
\begin{equation}
\label{eq:deltaK}
\delta K
=
\Omega_s
\left[
-\frac{\kappa}{\sqrt{2}}(r_++r_-)
+\frac{\kappa^2}{2}(r_++r_-)^2
+2\tau^2r_+r_-
\right].
\end{equation}
Thus, the leading geometry-independent contribution is simply
$\Omega_s$, whereas $\delta K$ contains the finite-curvature and
finite-torsion corrections.

We next introduce the circular momentum operators
$\hat k_\pm=(\hat k_1\mp i\hat k_2)/\sqrt{2}$, with
$\hat k_j=-i\partial_j$. They satisfy
$[\hat k_+,r_-]=[\hat k_-,r_+]=-i$ and
$[\hat k_+,r_+]=[\hat k_-,r_-]=0$.
In the helicity basis, the curl operator takes the form
\begin{equation}
\label{eq:curl_hel}
    \hat{\mathcal K}
    \equiv
    i[\hat{\mathbf k}\times]_{\mathrm{hel}}
    =
    \begin{pmatrix}
    \hat k_s & 0 & -\hat k_+
    \\[2mm]
    0 & -\hat k_s & \hat k_-
    \\[2mm]
    -\hat k_- & \hat k_+ & 0
    \end{pmatrix}.
\end{equation}
The rotation-induced Maxwell operator is therefore
$\hat{\mathcal H}_\Omega
\equiv i\hat{\mathcal K}\mathsf M_\Omega$.
In the Maxwell evolution equations, this operator contributes as
$-\hat{\mathcal H}_\Omega$. To display its structure compactly, we introduce
\begin{equation}
\label{eq:Qalphabeta}
    \hat{\mathcal Q}_{\alpha\beta}
    \equiv
    i\hat k_\alpha\!\left(Kr_\beta\,\right),
    \qquad
    \alpha,\beta=\pm ,
\end{equation}
where the dot indicates that $\hat k_\alpha$ acts on the product of
$Kr_\beta$ and the field to its right. The rotation-induced operator
can then be written as
\begin{equation}
\label{eq:Homega_full}
\hat{\mathcal H}_\Omega
=
\begin{pmatrix}
\hat k_sg_0+\hat{\mathcal Q}_{+-}
&
\hat{\mathcal Q}_{++}
&
i\hat k_s(Kr_+)
\\[2mm]
-\hat{\mathcal Q}_{--}
&
\hat k_sg_0-\hat{\mathcal Q}_{-+}
&
-i\hat k_s(Kr_-)
\\[2mm]
-\hat k_-g_0
&
-\hat k_+g_0
&
-\hat{\mathcal Q}_{-+}
+\hat{\mathcal Q}_{+-}
\end{pmatrix}.
\end{equation}
Because $K$ depends on the transverse coordinates, the transverse
momentum operators do not in general commute with it. In particular,
the combination appearing in the longitudinal component can be
decomposed as
\begin{equation}
\label{eq:ordering_identity}
i\hat k_+(Kr_-)
-i\hat k_-(Kr_+)
=
K\hat L_s
+
\Delta_{\mathrm{ord}},
\end{equation}
where
$\hat L_s
=i(r_-\hat k_+-r_+\hat k_-)$
is the OAM operator along the local propagation
direction, and
\begin{equation}
    \Delta_{\mathrm{ord}}
    =
    i\left[
    r_-(\hat k_+K)
    -
    r_+(\hat k_-K)
    \right]
    =
    [\hat L_s,K].
\end{equation}
Using Eq.~\eqref{eq:K_simplified}, the ordering correction for the
present axial rotation becomes
\begin{equation}
\label{eq:Delta_ord}
\Delta_{\mathrm{ord}}
=
\frac{\kappa\Omega_s}{\sqrt{2}}
(r_+-r_-)
\left[
1-\sqrt{2}\kappa(r_++r_-)
\right].
\end{equation}
Let $a_\perp$ denote the characteristic transverse size of the guided
mode. In the weakly curved and weakly twisted regime,
\begin{equation}
\label{eq:weak_geometry}
    \kappa a_\perp\ll1,
    \qquad
    \tau a_\perp\ll1,
\end{equation}
the geometry-dependent corrections are parametrically small,
\begin{equation}
    \frac{\delta K}{\Omega_s}
    =
    \mathcal O(\kappa a_\perp)
    +
    \mathcal O[(\kappa a_\perp)^2]
    +
    \mathcal O[(\tau a_\perp)^2],
    \,
    \frac{\Delta_{\mathrm{ord}}}{\Omega_s}
    =
    \mathcal O(\kappa a_\perp).
\end{equation}
Thus, to leading order in the weak-geometry expansion,
$K\simeq\Omega_s$, while both $\delta K$ and
$\Delta_{\rm ord}$ contribute only perturbative geometric corrections.
Moreover, since
$r_\pm=(\rho/\sqrt{2})e^{\mp i\phi}$, the terms linear in $r_\pm$
primarily couple neighboring OAM channels,
$l\rightarrow l\mp1$. For an approximately axisymmetric guided mode
with a well-defined OAM quantum number $l$, these
terms do not contribute to the leading diagonal rotation-induced phase
shift. They can therefore be consistently absorbed into the
perturbative part of the effective Hamiltonian.

We next derive the optical Dirac-like equation for the electromagnetic
field in the rotating helical waveguide. Using the effective
constitutive relations in Eq.~\eqref{eq:constitutive_rotation}, together
with
\begin{equation}
        \mathbf H
    \simeq
    \mathsf G\mathbf B
    -
    \mathsf M_\Omega\mathbf D,
    \qquad
    \mathbf E
    \simeq
    \mathsf G\mathbf D
    +
    \mathsf M_\Omega\mathbf B .
\end{equation}
Maxwell's equations take the form
\begin{equation}
\label{eq:Maxwell_rotating_medium}
\begin{aligned}
    \frac{\partial\mathbf D}{\partial t}
    &=
    \boldsymbol{\nabla}\times
    \left(
        \mathsf G\mathbf B
        -
        \mathsf M_\Omega\mathbf D
    \right),
    \\
    \frac{\partial\mathbf B}{\partial t}
    &=
    -
    \boldsymbol{\nabla}\times
    \left(
        \mathsf G\mathbf D
        +
        \mathsf M_\Omega\mathbf B
    \right).
\end{aligned}
\end{equation}
For either electromagnetic field,
$\mathbf V=\mathbf D$ or $\mathbf B$, we first evaluate the
rotation-induced contribution
$\boldsymbol{\nabla}\times(\mathsf M_\Omega\mathbf V)$.
Using the helicity-space representations of the curl operator and
$\mathsf M_\Omega$ given in Eqs.~(\ref{eq:Momega_def}) and (\ref{eq:Momega}), and eliminating the
longitudinal field component through the transversality condition, the
resulting transverse operator can be organized as
\begin{equation}
\label{eq:Homega_transverse}
\left(
\hat{\mathcal H}_\Omega\mathbf V
\right)_\perp
\simeq
\left[
    \hat{\mathcal H_0}\mathbf 1_2
    +
    K\hat J_s^{(2)}
    +
    \hat{\mathcal H}_{\mathcal Q}
    +
    \hat{\mathcal H}_{\rm ord}
\right]
\mathbf V_\perp ,
\end{equation}
where
$\hat J_s^{(2)}=\hat L_s\mathbf 1_2+\sigma_3$
is the total AM operator in the transverse two-component
space, and $H_0=g_0\hat k_s$. The operator $\hat{\mathcal V}_\perp$ describes the transverse
mode-mixing contribution and can be written as
\begin{equation}
    \hat{\mathcal H}_{\mathcal Q}= \begin{pmatrix}
        0&\hat{\mathcal Q}_{++}
        \\
        -\hat{\mathcal Q}_{--}&0
    \end{pmatrix},
\end{equation}
The remaining operator
$\hat{\mathcal H}_{\rm ord}$ collects the ordering corrections generated
by transverse derivatives acting on the coordinate-dependent coefficient
$K$, including the contribution defined in
Eq.~\eqref{eq:Delta_ord}. Using the decomposition $K=\Omega_s+\delta K$,
Eq.~\eqref{eq:Homega_transverse} becomes
\begin{equation}
\label{eq:Homega_split}
\left(
\hat{\mathcal H}_\Omega\mathbf V
\right)_\perp
\simeq
\left[
   \hat{\mathcal H_0}\mathbf 1_2
    +
    \Omega_s\hat J_s^{(2)}
    +
    \hat{\mathcal H_{\delta}}
\right]
\mathbf V_\perp ,
\end{equation}
where$
    \hat{\mathcal H_{\delta}}
    =
    \delta K\,\hat J_s^{(2)}
    +
    \hat{\mathcal H}_{\mathcal Q}
    +
    \hat{\mathcal H}_{\rm ord}$. Equation~\eqref{eq:Homega_split} therefore explicitly separates the
leading rotation--AM coupling
$\Omega_s\hat J_s^{(2)}$ from the geometry-dependent, mode-mixing, and
operator-ordering corrections.

In the absence of mechanical rotation, the intrinsic helical-waveguide
contribution follows from the same procedure applied to
$\boldsymbol{\nabla}\times(\mathsf G\mathbf V)$. After eliminating the
longitudinal component, the corresponding transverse Maxwell operator
takes the form\cite{yang2025inducedberryconnectionphotonic}
\begin{equation}
\label{eq:Hhel}
    \hat{\mathcal H}_{\rm hel}
    =
    \sigma_3
    \left[
        \hat k_s
        +
        \tau\hat J_s^{(2)}
    \right]
    +
    \hat{\mathcal H}_{1/k_s},
\end{equation}
where $\hat{\mathcal H}_{1/k_s}$ contains the higher-order paraxial
corrections,
\begin{equation}
\label{eq:H1ks}
    \hat{\mathcal H}_{1/k_s}
    =
    \frac{1}{\hat k_s}
    \left(
        \Pi_I\mathbf 1_2
        +
        \boldsymbol{\Pi}_\perp
        \cdot
        \boldsymbol{\sigma}_\perp
        +
        \Pi_3\sigma_3
    \right).
\end{equation}
The corresponding geometric coefficients are
\begin{equation}
    \begin{aligned}
         \Pi_I
    &=
    G_{ss}\hat k_\perp^2
    -
    i\boldsymbol{\eta}_\perp
    \cdot
    \hat{\mathbf k}_\perp,
    \\
    \Pi_\pm
    &=
    G_{ss}\hat k_\pm^2
    +
    2i
    \left(
        \tau\hat k_s r_\pm+\eta_\pm
    \right)
    \hat k_\pm,
    \\
    \Pi_3
    &=
    \left(
        \boldsymbol{\eta}_\perp
        \times
        \hat{\mathbf k}_\perp
    \right)_s .
    \end{aligned}
\end{equation}
Here, $\boldsymbol{\eta}_\perp=1/2 \boldsymbol{\nabla}_\perp G_{ss}$ and $\eta_\pm$ denote the corresponding circular components of
$\boldsymbol{\eta}_\perp$.

To expose the Dirac-like structure, we introduce the four-component
optical wave function
\begin{equation}
    \boldsymbol{\Psi}
    =
    \begin{pmatrix}
        \boldsymbol{\Psi}_+
        \\
        \boldsymbol{\Psi}_-
    \end{pmatrix},
    \qquad
    \boldsymbol{\Psi}_\pm
    =
    \mathbf D_\perp
    \pm
    i\sigma_3\mathbf B_\perp .
\end{equation}
Combining the intrinsic helical-waveguide contribution \eqref{eq:Hhel} with the
mechanical-rotation term \eqref{eq:Homega_split}, we obtain the effective four-component Dirac-like equation
\begin{equation}
\label{eq:final_wave}
i\frac{\partial\boldsymbol{\Psi}}{\partial t}
+
\Omega_s\hat J_s\boldsymbol{\Psi}
=
\left[
    \left(
        \gamma^0-g_0\mathbf 1_4
    \right)\hat k_s
    +
    \gamma^0\tau\hat J_s
\right]
\boldsymbol{\Psi}
+
\hat H'\boldsymbol{\Psi}.
\end{equation}
Here $\gamma^0$ is the Dirac branch matrix, which in the present
representation takes the form
\begin{equation}
     \gamma^0
    =
    \begin{pmatrix}
        \mathbf 1_2&0
        \\
        0&-\mathbf 1_2
    \end{pmatrix},
\end{equation}
so that the upper and lower two-component sectors correspond to the
$\gamma^0$ eigenvalues $+1$ and $-1$, respectively. More generally, the coupling between mechanical rotation and optical
AM takes the form $-\boldsymbol{\Omega}\cdot\hat{\mathbf J}$. In the present formulation, the spatial coordinate and the associated
rotation-induced vector $\mathbf g_\Omega$ have been separated into
extrinsic and intrinsic parts. As a result, the term displayed
explicitly in Eq.~\eqref{eq:final_wave},
$-\Omega_s\hat J_s$, represents the leading longitudinal coupling
between the mechanical rotation and the intrinsic AM of
the guided mode. The remaining rotation-induced contributions,
including those associated with the extrinsic orbital motion, are
absorbed into the residual operator $\hat H'$ together with the
higher-order geometric and paraxial corrections. The remaining
subleading contributions are collected in
\begin{equation}
    \hat H'
    =
    \hat{\mathcal H}_{1/k_s}^{(4)}
    -
    \hat{\mathcal H_{\delta}}^{(4)},
\end{equation}
where the superscript $(4)$ denotes the corresponding extension of the
two-component transverse operators to the four-component space. In this representation, the spin and total AM operators
along the local propagation direction are
\begin{equation}
    \hat S_s
    =
    \begin{pmatrix}
        \sigma_3&0
        \\
        0&\sigma_3
    \end{pmatrix},
    \qquad
    \hat J_s
    =
    \hat L_s\mathbf 1_4+\hat S_s .
\end{equation}
Equation~\eqref{eq:final_wave} makes explicit the two leading
AM-dependent interactions,
\begin{equation}
\label{eq:leading_interactions}
    \hat H_\tau
    =
    \gamma^0\tau\hat J_s,
    \qquad
    \hat H_\Omega
    =
    -\Omega_s\hat J_s .
\end{equation}
The first originates from the intrinsic torsion of the helical
waveguide, whereas the second represents the coupling of the optical
total AM to the mechanical rotation of the fiber. The
remaining curvature-, torsion-, transverse-mode-mixing, and
operator-ordering corrections are perturbative in the weak-geometry
regime
$\kappa a_\perp\ll1$ and $\tau a_\perp\ll1$.

\section{Derivation of the Conventional Sagnac Phase from Extrinsic Orbital Angular Momentum}\label{appendix3}
The conventional area-dependent Sagnac phase can be understood as the
phase generated by the extrinsic part of the rotation--AM
coupling,
\begin{equation}
\label{eq:H_ext_rotation}
    \hat H_{\Omega}^{\rm ext}
    =
    -\boldsymbol{\Omega}\cdot\hat{\mathbf L}_{\rm ext},
    \qquad
    \hat{\mathbf L}_{\rm ext}
    =
    \mathbf r\times\hat{\mathbf p},
\end{equation}
Using
$\boldsymbol{\Omega}\cdot(\mathbf r\times\hat{\mathbf p})
=(\boldsymbol{\Omega}\times\mathbf r)\cdot\hat{\mathbf p}$,
the interaction may equivalently be written as
$\hat H_{\Omega}^{\rm ext}
=-(\boldsymbol{\Omega}\times\mathbf r)\cdot\hat{\mathbf p}$.

In the semiclassical propagation picture, the corresponding phase is
$\phi_{\rm ext}=-\int H_{\Omega}^{\rm ext}\,dt$.
Under the idealized Sagnac approximation adopted here, in which
medium-drag and higher-order dispersive corrections are neglected, this
reduces to
\begin{equation}
\label{eq:external_single_pass}
    \phi_{\rm ext}^{(+)}
    =
    \frac{\omega}{c^2}
    \int_{\rm helix}
    \left(
    \boldsymbol{\Omega}\times\mathbf r
    \right)\cdot d\mathbf r .
\end{equation}
Thus, the conventional Sagnac phase can be viewed directly as the
semiclassical manifestation of the extrinsic coupling
$-\boldsymbol{\Omega}\cdot\hat{\mathbf L}_{\rm ext}$.

For a uniform helix parameterized by
$\mathbf r(\varphi)
=(R\cos\varphi,R\sin\varphi,p_h\varphi)$ and rotating about its axis
with $\boldsymbol{\Omega}=\Omega_z\mathbf e_z$, one finds
$(\boldsymbol{\Omega}\times\mathbf r)\cdot d\mathbf r
=\Omega_zR^2\,d\varphi$. Hence, propagation over $N$ turns gives the
single-pass phase
\begin{equation}
    \phi_{\rm ext}^{(+)}
=2\pi N\omega\Omega_zR^2/c^2.
\end{equation}

For a conventional Sagnac interferometer, two counterpropagating
beams traverse the closed optical path in opposite directions
and are subsequently recombined at the output. Here, $(+)$
denotes propagation along the helical segment in the direction
of increasing $\varphi$, whereas $(-)$ denotes the opposite
direction. Reversing the propagation direction reverses the
orientation of the line integral, so that
$\phi_{\rm ext}^{(-)}=-\phi_{\rm ext}^{(+)}$.
The measured phase difference is therefore
\begin{equation}
\label{eq:Sagnac_external}
    \Delta\Phi_{\rm ext}
    =
    \phi_{\rm ext}^{(+)}
    -
    \phi_{\rm ext}^{(-)}
    =
    \frac{4\pi N\omega\Omega_zR^2}{c^2}
    =
    \frac{4\omega\Omega_z}{c^2}A_{\rm helix}.
\end{equation}
where $A_{\rm helix}=N\pi R^2$ is the projected oriented area associated
with $N$ turns of the helical trajectory.

\section{Mirror reflection and transformation of intrinsic
angular momentum}
\label{app:mirror_reflection}

We derive the transformation of the optical Dirac wavefunction
and its intrinsic AM under mirror reflection.
Both the incident and reflected fields are initially expressed
in the same fixed Frenet frame
$(\mathbf e_1,\mathbf e_2,\mathbf e_s)$.
We consider ideal, locally normal, mode-preserving reflection
in the paraxial limit.

For a perfectly conducting mirror at $s=L$, the tangential
electric field vanishes at the reflecting surface,
\begin{equation}
    \mathbf e_s\times
    \left(
        \mathbf E^{(i)}+\mathbf E^{(r)}
    \right)=0.
\end{equation}
Thus, $\mathbf E_\perp^{(r)}=-\mathbf E_\perp^{(i)}$.
From Faraday's law,
$\mathbf B=(\mathbf k\times\mathbf E)/\omega$,
and the reversal of the propagation direction
$\mathbf k_r=-\mathbf k_i$, we obtain
$\mathbf B_\perp^{(r)}=\mathbf B_\perp^{(i)}$.
For the locally transverse, mode-preserving approximation,
the displacement field also satisfies
$\mathbf D_\perp^{(r)}\simeq-\mathbf D_\perp^{(i)}$. Using the optical Dirac wavefunction $\boldsymbol{\Psi}_\pm$ defined in the main text, the reflection conditions give
\begin{equation}
\begin{aligned}
    \boldsymbol{\Psi}_+^{(r)}
    \simeq-\boldsymbol{\Psi}_-^{(i)},\quad \boldsymbol{\Psi}_-^{(r)}
    \simeq-\boldsymbol{\Psi}_+^{(i)}.
\end{aligned}
\end{equation}
Hence, the reflection operator in the fixed Frenet
representation is
\begin{equation}
\label{eq:mirror_operator_appendix}
    \boldsymbol{\Psi}^{(r)}
    =
    \hat R_{\rm mirror}\boldsymbol{\Psi}^{(i)},
    \qquad
    \hat R_{\rm mirror}
    =
    -
    \begin{pmatrix}
        0&I_2\\
        I_2&0
    \end{pmatrix}.
\end{equation}
The reflection operator is obtained in the leading
paraxial, mode-preserving approximation. In the rotating
effective medium, the magnetoelectric coupling may
introduce first-order corrections to the transverse
constitutive fields and hence to the reflection amplitudes.
For a locally isotropic, impedance-matched medium, these
corrections give a real reflection-amplitude factor and
do not introduce an additional first-order reflection
phase. We therefore neglect the corresponding amplitude
corrections and assume that the end mirror does not induce
additional mode mixing or AM-dependent
reflection phases at the order considered.

Since $\gamma^0=\mathrm{diag}(I_2,-I_2)$,
the reflection operator satisfies
\begin{equation}
\label{eq:mirror_gamma_appendix}
    \hat R_{\rm mirror}^{-1}
    \gamma^0
    \hat R_{\rm mirror}
    =
    -\gamma^0.
\end{equation}
Therefore, reflection exchanges the forward- and
backward-propagating branches,
$\boldsymbol{\Psi}_+\rightarrow\boldsymbol{\Psi}_-$.
Both branches correspond to positive physical frequencies,
since reversing the eigenvalue of $\gamma^0$ is accompanied
by the reversal of the longitudinal wave vector.

The intrinsic total AM operator
along the fixed Frenet axis is
$\hat J_s=\hat L_sI_4+\hat S_s$, as defined in the main text.
Because $\hat R_{\rm mirror}$ is independent of the transverse
coordinates and acts only on the propagation-branch indices,
it commutes with $\hat J_s$. Consequently,
\begin{equation}
\label{eq:mirror_J_appendix}
    \hat R_{\rm mirror}^{-1}
    \hat J_s
    \hat R_{\rm mirror}
    =
    \hat J_s.
\end{equation}
Thus, the eigenvalue $j=l+\sigma$ projected onto
the original fixed Frenet axis is preserved under reflection.

For the propagation-referenced OAM and SAM operators,
$\hat L_{\parallel}=\gamma^0\hat L_sI_4$
and $\hat S_{\parallel}=\gamma^0\hat S_s$,
the reflection transformation yields
\begin{equation}
\begin{aligned}
    \hat R_{\rm mirror}^{-1}
    \hat L_{\parallel}
    \hat R_{\rm mirror}
    =-\hat L_{\parallel},\quad \hat R_{\rm mirror}^{-1}
    \hat S_{\parallel}
    \hat R_{\rm mirror}
    =-\hat S_{\parallel}.
\end{aligned}
\end{equation}
Consequently, the propagation-referenced total
AM operator
$\hat J_{\parallel}=\hat L_{\parallel}+\hat S_{\parallel}$
satisfies
\begin{equation}
\label{eq:mirror_J_parallel}
    \hat R_{\rm mirror}^{-1}
    \hat J_{\parallel}
    \hat R_{\rm mirror}
    =
    -\hat J_{\parallel}.
\end{equation}
For an incident mode satisfying
$\hat J_{\parallel}\boldsymbol{\Psi}_+
=j\boldsymbol{\Psi}_+$, the reflected mode
$\boldsymbol{\Psi}_-=\hat R_{\rm mirror}\boldsymbol{\Psi}_+$
therefore satisfies
\begin{equation}
    \hat J_{\parallel}\boldsymbol{\Psi}_-
    =
    -j\boldsymbol{\Psi}_-.
\end{equation}
This transformation is consistent with the
reversal of the propagation-referenced orbital
angular momentum of reflected optical vortex beams,
as discussed by Bliokh and Aiello
\cite{BliokhAiello2013}.
For the ideal metallic mirror considered here,
the helicity also reverses relative to the
propagation direction, while the spin angular
momentum projected onto the fixed Frenet axis
is preserved. Accordingly, the OAM and SAM
quantum numbers referred to the propagation direction
transform as
\begin{equation}
l\longrightarrow-l,
    \qquad
    \sigma\longrightarrow-\sigma,
    \qquad
    j=l+\sigma\longrightarrow-j.
\end{equation}
This result is consistent with the conservation of
$\hat J_s$ in the fixed Frenet representation:
mirror reflection preserves the AM
projection onto the original axis but reverses its
projection onto the actual propagation direction.

Finally, the two AM-dependent interactions,
$\hat H_\tau=\gamma^0\tau\hat J_s$ and
$\hat H_\Omega=-\Omega_s\hat J_s$,
transform differently under reflection:
\begin{equation}
\begin{aligned}
    \hat R_{\rm mirror}^{-1}
    \hat H_\tau
    \hat R_{\rm mirror}
    &=-\hat H_\tau,\\
    \hat R_{\rm mirror}^{-1}
    \hat H_\Omega
    \hat R_{\rm mirror}
    &=\hat H_\Omega.
\end{aligned}
\end{equation}
Therefore, the torsion-induced geometric phases
cancel between the two propagation directions,
whereas the intrinsic mechanical-rotation phases
retain the same sign and accumulate constructively,
as derived in the main text.

\bibliography{main}

\end{document}